\documentclass[aps,prb,reprint,superscriptaddress,floatfix]{revtex4-2}
\usepackage{amsmath,amssymb,bm,booktabs,graphicx}
\usepackage{pgfplots}
\pgfplotsset{compat=1.18}
\usepackage[colorlinks=true,allcolors=blue]{hyperref}
\providecommand{\orcid}[1]{}
\graphicspath{{figs/}}
\newcommand{\HH}{\mathbb H}
\newcommand{\GI}{\mathrm{GI}}
\newcommand{\Cdis}{\mathrm{C}}
\newcommand{\Expdis}{\mathrm{Exp}}
\begin{document}
\title{Covariance-dependent participation scaling of Comtet disorder on open hyperbolic lattices}
\author{Wei\ Li}
\orcid{0000-0003-2586-6057}
\email{wliustc@aust.edu.cn}
\affiliation{Center for Fundamental Physics, Anhui University of Science and Technology, Huainan 232001, China}
\author{Yong\ Cheng}
\affiliation{School of Mechanics and Optoelectrical Physics, Anhui University of Science and Technology, Huainan, Anhui 232001, People's Republic of China}
\date{October 31, 2026}
\begin{abstract}
Negative curvature makes the electrostatic Green function decay exponentially, producing a finite disorder-variance budget even as radial volume grows exponentially. We investigate whether this near-field concentration also determines wave-function participation on open hyperbolic lattices. A Gaussian control preserves the full one-site distribution of Comtet disorder at every vertex while removing its off-diagonal covariance. Complete-window participation ratios on $\{7,3\}$ graphs with $N=847,2240,5887$ yield an effective-exponent difference of $-0.028$ at $W/t=0.5$ and $+0.116$ at $W/t=3$; both signs persist upon extending to $N=15435$. Size-resolved squared-kernel radii lie near $2.45\ell$ on the larger graphs of this sequence, below the cutoff-matched continuum value $3.12\ell$. Distance-shell weights resolve the finite discrete measure associated with this shift, while higher participation moments and spacing ratios characterize the accompanying wave-function and spectral structure. A marginal-matched exponential-kernel family gives positive exponent contrasts at all five sampled kernel lengths at $W/t=0.5,3$, separating the Comtet response from a description based on the decay length alone. These results distinguish a concentrated variance budget from the covariance-dependent participation response of finite open hyperbolic systems.
\end{abstract}
\maketitle
\section{Introduction}
\label{sec:intro}

Spatial correlations in a random potential influence how a quantum state samples disorder. The potential distribution at a single site fixes the local disorder scale, while correlations specify how different sites fluctuate together. Separating these contributions is useful whenever the disorder originates from a spatial kernel rather than independent onsite variables. Localization theory provides the broader setting for this distinction \cite{Anderson1958,Abrahams1979,LeeRamakrishnan1985,EversMirlin2008}; correlated-disorder models show how spatial organization can modify localization even at fixed local disorder scales \cite{dunlap1990Absence,deMouraLyra1998,IzrailevKrokhin1999,Titov2005,izrailev2012Anomalous}. On a curved lattice, the spatial kernel and the measure used to sum its contributions are both set by geometry.

Hyperbolic lattices combine local connectivity with exponential growth of the surrounding volume. Their geometry has motivated hyperbolic crystallography \cite{boettcher2022Crystallography}, extensions of Bloch theory \cite{MaciejkoRayan2021,Maciejko2022}, and descriptions incorporating higher-dimensional representations and non-Abelian supercells \cite{cheng2022Band,Lenggenhager2023}. Topological band models \cite{Yu2020,Urwyler2022} and circuit and photonic platforms \cite{Kollar2019,Lenggenhager2022,Chen2023HC,Park2024} provide concrete settings in which this geometry shapes spectral and wave-function structure. Disorder on such lattices raises questions that differ from those on a Euclidean square lattice, because the number of sites in successive shells grows rapidly and an open cluster retains a substantial boundary population. Studies of Anderson localization on hyperbolic lattices illustrate the role of graph geometry and boundary construction \cite{Chen2024AL,Li2024}. Continuum weak-localization theory on negatively curved surfaces \cite{Curtis2025}, charge-disorder studies of hyperbolic Dirac liquids \cite{Leong2025}, and mass-disorder models \cite{Sen2026MassDisorder} address complementary observables and disorder operators. Here we study a spatially correlated scalar potential on a fixed sequence of open graphs and resolve its effect through a control with identical local Gaussian distributions.

Electrostatic disorder is also a familiar source of spatially structured potentials in planar materials, including graphene \cite{Adam2007,Peres2010,DasSarma2011}. On a hyperbolic graph, relating such a potential to continuum geometry requires specifying both distances and the discrete measure \cite{boettcher2020Quantum}. The electrostatic Green function on the hyperbolic plane supplies a natural kernel for this problem. With curvature $-1/\ell^2$, it has a logarithmic singularity near its source and decays as $e^{-r/\ell}$ far away. The squared kernel is integrable against the radial area measure, which grows as $e^{r/\ell}$. This balance makes the variance of a field generated by independent Gaussian sources finite in the continuum and strongly concentrated at short distances on the finite graphs considered here. A short variance radius, however, specifies how source contributions build the local second moment. The spatial covariance of the resulting potential depends on the overlap of two kernels. These are distinct objects, and their effects on a wave function must be assessed separately.

We make that separation exact at the ensemble level. The Comtet potential is a linear transformation of independent Gaussian sources. Its independent-site Gaussian control has the same variance at every vertex, and hence the same complete Gaussian marginal distribution, while setting all off-diagonal covariances to zero. Boundary-dependent local variances are retained in both ensembles. The contrast therefore measures the response to covariance on a common graph, without changing the local potential distributions.

Our main result is a disorder-dependent sign contrast in finite-size participation scaling. In a complete energy window around zero, the Comtet potential produces a smaller effective participation exponent than the Gaussian control at $W/t=0.5$, and a larger exponent at $W/t=3$. An additional graph with $15435$ sites reproduces both signs. The intermediate point, $W/t=1$, has an unresolved exponent difference. We complement this comparison with higher participation moments, adjacent-level spacing ratios, and a family of row-normalized exponential kernels that preserves the same local Gaussian distributions. The exponential family yields a positive contrast at every sampled kernel length in both tested disorder regimes.

A separate geometric analysis resolves the variance budget. The radius enclosing $95\%$ of the squared-kernel weight changes little along the largest available graph sizes, yet lies below the corresponding continuum radius. Comparing the discrete distance measure with the continuum shell measure identifies how near shells and the finite graph's pair distribution enter this shift. These observations organize the connection between geometric disorder construction and wave-function response without replacing either observable by a single effective length.

Section~\ref{sec:model} defines the graph, disorder ensembles, and averaging conventions. Section~\ref{sec:variance} presents the variance profiles and their continuum comparison. Sections~\ref{sec:participation}--\ref{sec:expresults} give participation scaling, spectral diagnostics, and the exponential-kernel comparison. Section~\ref{sec:discussion} discusses their physical interpretation. The appendices provide the Green-function derivation, shell and support controls, statistical details, and the geometric constraints on compact boundary conditions.
\section{Model and observables}
\label{sec:model}

\subsection{Open hyperbolic graphs}

We consider vertex graphs of regular hyperbolic tilings $\{p,q\}$, with $p$ edges per polygon and $q$ polygons meeting at a bulk vertex. The hyperbolic condition is $(p-2)(q-2)>4$. Finite patches are generated in the Poincar\'e disk with \texttt{hypertiling} \cite{hypertiling}. The main sequence is $\{7,3\}$ with layer labels $n=5,6,7,8$ and $N=847,2240,5887,15435$, respectively. The $\{8,3\}$ variance sequence uses $n=5,6$ with $N=2888,10800$. Both tilings have bulk coordination three. All patches have open boundaries.

For disk coordinates $z_i$ with $|z_i|<1$, the geodesic distance is
\begin{equation}
d_{ij}=\ell\,\operatorname{arcosh}\!\left[
1+\frac{2|z_i-z_j|^2}{(1-|z_i|^2)(1-|z_j|^2)}
\right].
\label{eq:distance}
\end{equation}
The single-particle Hamiltonian is
\begin{equation}
H=-t\sum_{\langle i,j\rangle}
\left(|i\rangle\langle j|+|j\rangle\langle i|\right)
+\sum_i V_i|i\rangle\langle i|.
\label{eq:H}
\end{equation}
An edge is included once in the undirected nearest-neighbor sum. Energies are measured in units of $t$ and lengths in units of $\ell$. The graph boundary is part of the model and of the observables below.

\subsection{Comtet field and exact marginal matching}

The radial electrostatic kernel on $\HH^2$ is
\begin{equation}
\Phi_H(r)=\frac{1}{2\pi}\ln\coth\frac{r}{2\ell}.
\label{eq:Phi}
\end{equation}
Appendix~\ref{app:green} derives its normalization and asymptotic forms. On each graph we define the symmetric matrix
\begin{equation}
K_{ij}=\mathbf 1_{i\ne j}\mathbf 1_{d_{ij}\le R_C}\Phi_H(d_{ij}),
\qquad R_C=6\ell,
\label{eq:K}
\end{equation}
excluding the singular self-term. With
\begin{equation}
S_2=\frac{1}{N}\sum_{i,j}K_{ij}^2,\qquad
a_W=\frac{W}{\sqrt{12S_2}},
\label{eq:S2}
\end{equation}
the correlated field is
\begin{equation}
V_i^\Cdis=a_W\sum_jK_{ij}\eta_j,\qquad
\eta_j\overset{\mathrm{iid}}{\sim}{\cal N}(0,1).
\label{eq:VC}
\end{equation}
It has covariance
\begin{equation}
\Sigma^\Cdis_{ik}
=\mathbb E[V_i^\Cdis V_k^\Cdis]
=a_W^2\sum_jK_{ij}K_{kj}.
\label{eq:cov}
\end{equation}
Thus $\sigma_i^2=\Sigma^\Cdis_{ii}=a_W^2\sum_jK_{ij}^2$ and $N^{-1}\sum_i\sigma_i^2=W^2/12$. The normalization sets the site-averaged ensemble variance while retaining the graph-dependent variance at each vertex.

The Gaussian independent-site control, denoted GI, is
\begin{equation}
V_i^\GI=\sigma_i\zeta_i,\qquad
\zeta_i\overset{\mathrm{iid}}{\sim}{\cal N}(0,1),
\label{eq:GI}
\end{equation}
with independent source streams for the two ensembles. At every site,
\begin{equation}
V_i^\Cdis\sim{\cal N}(0,\sigma_i^2),\qquad
V_i^\GI\sim{\cal N}(0,\sigma_i^2).
\end{equation}
Their full one-site distributions coincide, including the spatial variation of the variance induced by the boundary. Their difference is the off-diagonal covariance:
$\Sigma^\GI_{ik}=\delta_{ik}\sigma_i^2$.
For Gaussian fields, the covariance matrix determines the full joint distribution. Equations~\eqref{eq:cov} and \eqref{eq:GI} therefore specify precisely which ensemble information is changed.

The discrete $S_2$ is evaluated separately for every graph. The values are $0.274$--$0.277$ for the first three $\{7,3\}$ sizes and approximately $0.137$ for $\{8,3\}$. These sums include the graph's actual vertex density and boundary geometry. The continuum radial integral derived below provides a separate geometric reference.

\subsection{Variance shells and quantile radius}

The normalized discrete squared-kernel measure is
\begin{equation}
\begin{aligned}
d\mu_N(r)&=\frac{\sum_{i,j}K_{ij}^2
\,\delta(r-d_{ij})\,dr}{NS_2},\\
F_N(r)&=\int_0^r d\mu_N(s).
\end{aligned}
\label{eq:measure}
\end{equation}
All ordered pairs with $i\ne j$ enter. The radius $r_{95}$ is the $95\%$ quantile of this measure, evaluated with a specified histogram and interpolation rule. The size-resolved analysis uses bins of width $h=0.05\ell$, right-open bins, and linear interpolation between neighboring bin centers. We keep this estimator fixed along the size sequence. The bin-step sensitivity is $u_{\rm disc}=|r_{95}(h/2)-r_{95}(h)|$.

For comparison we use
\begin{equation}
F_R^{\rm ct}(r)=
\frac{\int_0^{\min(r,R)}\Phi_H^2(s)\sinh(s/\ell)\,ds}
{\int_0^R\Phi_H^2(s)\sinh(s/\ell)\,ds}.
\label{eq:Fct}
\end{equation}
Constant angular and length factors cancel from this ratio. The reference with $R=6\ell$ shares the Comtet cutoff; $R=\infty$ defines a different reference. The squared-kernel radius measures the distances of sources contributing to a local variance, averaged over the graph. The potential covariance instead involves the two-kernel overlap in Eq.~\eqref{eq:cov}.

\subsection{Participation moments and spectral statistics}

For a normalized eigenstate, define
\begin{equation}
P_{q,\nu}=\sum_i|\psi_\nu(i)|^{2q}.
\end{equation}
The inverse participation ratio (IPR) is $P_{2,\nu}$. Every eigenstate in the complete window $\mathcal W_s=\{\nu:|E_{\nu,s}|<0.6t\}$ contributes. For realization $s$ with $m_s$ window states,
\begin{align}
I_s&=\frac{1}{m_s}\sum_{\nu\in{\cal W}_s}P_{2,\nu},\\
\overline I_N&=\frac{1}{S}\sum_{s=1}^S I_s,\qquad
r_N=\overline I_N^{-1}.
\label{eq:estimator}
\end{align}
The window includes all sites and all eigenstates satisfying the energy condition. Each realization receives equal weight, including when its number of window states differs. Equation~\eqref{eq:estimator} takes the inverse after both averages.

We fit $\ln r_N$ against $\ln N$ over $N=847,2240,5887$ to obtain the effective exponent $\alpha$. We denote the ensemble contrast by $\Delta\alpha=\alpha_\Cdis-\alpha_\GI$. An adjacent-size estimate is
\begin{equation}
\alpha_{ab}=
\frac{\ln r_{N_b}-\ln r_{N_a}}{\ln N_b-\ln N_a}.
\label{eq:pair}
\end{equation}
For general $q$, $\langle P_q\rangle$ uses the same within-realization state average followed by an equal-weight average over realizations. With this convention,
\begin{equation}
D_q=-\frac{1}{q-1}
\operatorname{slope}_{\ln N}\ln\langle P_q\rangle,
\qquad D_2=\alpha.
\label{eq:Dq}
\end{equation}
Arithmetic moment averaging defines an ensemble-averaged spectrum; typical averaging uses a different order of logarithm and ensemble average \cite{rodriguez2008Multifractal,vasquez2008Multifractal}. These dimensions use site number as the scaling variable. A uniformly extended state has $D_q=1$, while size-independent participation has $D_q=0$. Their numerical values are effective properties of this finite graph sequence.

Adjacent-gap ratios provide a spectral statistic that avoids unfolding \cite{OganesyanHuse2007,Atas2013}. For ordered window eigenvalues we also calculate adjacent gaps $\delta_\nu=E_{\nu+1}-E_\nu$ and
\begin{equation}
\rho_\nu=\frac{\min(\delta_\nu,\delta_{\nu+1})}
{\max(\delta_\nu,\delta_{\nu+1})}.
\label{eq:spacing}
\end{equation}
We first average $\rho_\nu$ within each realization. The reported summary then pools the per-realization means across the three main sizes within each disorder class and strength. It is a summary across graph sizes.

\subsection{Sampling and spectral windows}

The main C--GI data use $100$ realizations per class, disorder strength, and size. The three disorder strengths are $W/t=0.5,1,3$. The largest-cluster extension uses the same sample count at $W/t=0.5,3$. The principal exponent intervals are pointwise percentile bootstrap intervals, recomputing the fit after resampling independently within each class and size \cite{Efron1979}. The three-size table uses $300$ resamples; the adjacent-size table uses $10^4$. Sampling intervals and finite-size drift are reported as separate information.

The calculation uses dense diagonalization and shift-invert Lanczos for spectral windows \cite{Lanczos1950,ARPACK1998,SciPy2020}. The main three sizes contain $147$--$184$, $404$--$464$, and $1076$--$1197$ states per realization, respectively, with no empty windows. The resulting observable is a complete-window average, so changes in the spectral population are part of the defined comparison. Solver and cutoff checks are detailed in Appendix~\ref{app:numerics}.
\section{Variance concentration and distance-shell geometry}
\label{sec:variance}

\subsection{Near-field variance budget}

\begin{figure}[tb]
\includegraphics[width=\linewidth]{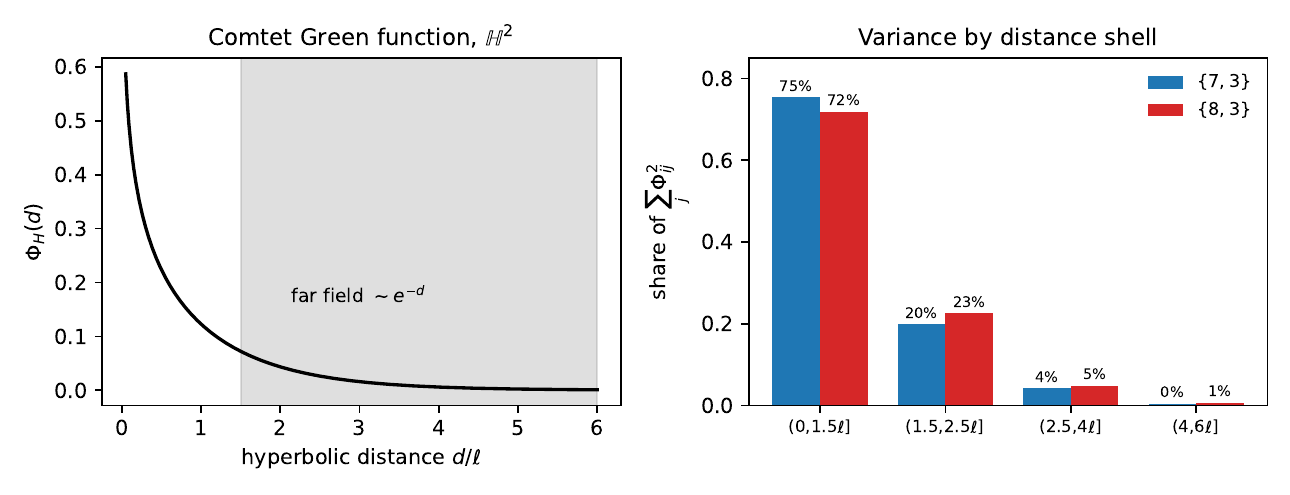}
\caption{Comtet kernel and its broad-shell variance decomposition on the indicated open hyperbolic graphs. Distance is the geodesic site separation. The shell weights refer to the source-distance measure of Eq.~\eqref{eq:measure}. They quantify the local-variance budget, while the field covariance is the two-kernel overlap of Eq.~\eqref{eq:cov}.}
\label{fig:kernel}
\end{figure}

The squared Comtet kernel has most of its weight within a few curvature lengths. Figure~\ref{fig:kernel} and Table~\ref{tab:shell} resolve this statement into broad distance intervals. On the two $\{7,3\}$ graphs, approximately three quarters of the variance comes from $(0,1.5\ell]$ and another fifth from $(1.5\ell,2.5\ell]$. The $\{8,3\}$ graph has a slightly smaller first-interval share, compensated mainly by the next interval. Only $0.4\%$--$0.6\%$ lies beyond $4\ell$ in these examples. This concentration survives the growth in the number of distant pairs because their squared-kernel weights decay rapidly.

\begin{table}[tb]
\caption{Squared-kernel variance shares in broad distance intervals. Entries are percentages; rounding need not give an exact total of $100\%$.}
\label{tab:shell}
\centering
\begin{tabular}{lrrrr}
\toprule
Graph & $(0,1.5]$ & $(1.5,2.5]$ & $(2.5,4]$ & $(4,6]$\\
\midrule
$\{7,3\}$, $847$ &75.7&19.7&4.1&0.4\\
$\{7,3\}$, $5887$ &75.4&19.9&4.3&0.5\\
$\{8,3\}$, $10800$ &71.9&22.6&4.9&0.6\\
\bottomrule
\end{tabular}
\end{table}

The continuum comparison makes the balance of decay and growth explicit. The large-$r$ squared kernel behaves as $e^{-2r/\ell}$, whereas the radial measure behaves as $e^{r/\ell}$. Their product therefore has an integrable $e^{-r/\ell}$ tail. The full radial integral is
\begin{equation}
\int_0^\infty\Phi_H^2(r)\sinh(r/\ell)\,dr=\frac{\ell}{24},
\label{eq:integral}
\end{equation}
as derived in Appendix~\ref{app:green}. Near-field concentration follows from this balance, but its quantitative radius depends on the spatial measure used to sum the kernel.

\subsection{Size-resolved variance radii}

Table~\ref{tab:r95} fixes the binning and interpolation rule of Sec.~\ref{sec:model} and follows $r_{95}$ along each tiling separately. On $\{7,3\}$, the values increase from $2.4302\ell$ to $2.4499\ell$ as the site count grows from $847$ to $15435$. The successive changes become smaller. The two $\{8,3\}$ values are $2.5509\ell$ and $2.5524\ell$. These sequences establish that the squared-kernel quantile changes slowly over the available larger clusters.

\begin{table}[tb]
\caption{Size-resolved $95\%$ radii using $h=0.05\ell$. The last column is the bin-step sensitivity measure, not a disorder-sampling confidence interval. Dashes indicate no value quoted here.}
\label{tab:r95}
\centering
\begin{tabular}{ccrrr}
\toprule
Tiling & $n$ & $N$ & $r_{95}/\ell$ & $u_{\rm disc}/\ell$\\
\midrule
$\{7,3\}$&5&847&2.4302&0.0349\\
&6&2240&2.4432&0.0284\\
&7&5887&2.4481&0.0260\\
&8&15435&2.4499&0.0250\\
$\{8,3\}$&5&2888&2.5509&---\\
&6&10800&2.5524&---\\
\bottomrule
\end{tabular}
\end{table}

A cross-tiling survey gives $r_{95}=2.72\pm0.29\ell$, with one graph size per tiling and the spread defined as the population standard deviation (Appendix~\ref{app:survey}). This survey uses a different quantile convention from the fixed-estimator sequences in Table~\ref{tab:r95}; its spread characterizes geometric diversity.

The cutoff-matched continuum reference gives $r_{95,6\ell}^{\rm ct}=3.1228\ell$, while its infinite-support counterpart gives $3.1792\ell$. Relative to the first reference, the discrete radii have offsets of $0.673$--$0.693\ell$ on $\{7,3\}$ and $0.570$--$0.572\ell$ on $\{8,3\}$. A slowly varying discrete variance radius thus coexists with a persistent offset from the continuum reference.

\subsection{Why the discrete measure matters}

On a continuum plane, arbitrarily small source distances contribute with the radial measure $\sinh(r/\ell)\,dr$. On the graph, distinct vertices have a minimum separation and the pair-distance distribution contains discrete shells. The finite open patch also changes the relative number of far-separated pairs. All three ingredients enter the normalized measure in Eq.~\eqref{eq:measure}.

The $\{7,3\}$ graph with $847$ vertices illustrates this explicitly. Its first distinct-distance shell, at $0.5663\ell$, carries a fraction $0.3901$ of the squared-kernel weight. The shells at $0.9928\ell$ and $1.2136\ell$ add $0.2339$ and $0.1330$. These three shells already account for $0.7569$. Farther out, the cumulative fraction is $0.9379$ at $2.1916\ell$ and $0.9495$ at $2.3319\ell$ (Table~\ref{tab:distinct}). The latter is still below $0.95$; the interpolated quantile is determined by the full prescribed histogram.

A continuum cutoff at $6\ell$ leaves a substantial separation from the discrete radii. Appending a continuum tail to the discrete profile shifts its quantile, which remains below the infinite-support continuum reference (Appendix~\ref{app:offset}). The associated leading quadrature estimate is $2.08\times10^{-4}\ell$, far smaller than the observed discrete--continuum separation. These comparisons motivate resolving the full distance measure. The measured near-, intermediate-, and far-distance relative weights, $1.11$, $0.94$, and $0.34$, are consistent with the far shells carrying less normalized weight on the finite graph.

The minimum separation, shell populations, and normalization jointly shape the squared-kernel measure. The larger nearest-neighbor spacing and smaller offset of $\{8,3\}$ are also consistent with this geometric interpretation; the two tilings differ in several geometric features, so this comparison does not isolate a unique cause.

\section{Covariance-dependent participation scaling}
\label{sec:participation}

\subsection{A sign contrast at fixed Gaussian marginals}

The C--GI comparison reveals a wave-function response that the short variance radius alone does not determine. Table~\ref{tab:alpha} gives three-size exponents for the complete energy window. Both ensembles have smaller $\alpha$ at stronger disorder. Their difference, however, is negative at $W/t=0.5$ and positive at $W/t=3$.

\begin{table}[tb]
\caption{Three-size effective participation exponents, with $100$ realizations per class and state point. Intervals are pointwise $95\%$ percentile bootstrap intervals. Exponents and contrasts are rounded separately from their underlying estimates.}
\label{tab:alpha}
\centering
\begin{tabular}{rrrrl}
\toprule
$W/t$&$\alpha_\Cdis$&$\alpha_\GI$&$\Delta\alpha$&$95\%$ CI\\
\midrule
0.5&0.531&0.560&$-0.028$&$[-0.038,-0.018]$\\
1&0.493&0.492&$+0.002$&$[-0.016,+0.017]$\\
3&0.356&0.240&$+0.116$&$[+0.096,+0.136]$\\
\bottomrule
\end{tabular}
\end{table}

\begin{figure}[tb]
\centering
\begin{tikzpicture}
\begin{axis}[width=8.4cm,height=5.2cm,xlabel={$W/t$},
ylabel={$\alpha_\Cdis-\alpha_\GI$},xmin=0.2,xmax=3.25,
ymin=-0.085,ymax=0.165,xtick={0.5,1,3},
tick label style={font=\small},label style={font=\small},
legend style={font=\scriptsize,at={(0.04,0.96)},anchor=north west},
legend cell align=left]
\addplot[gray,densely dashed,forget plot] coordinates {(0.2,0)(3.25,0)};
\addplot[blue!70!black,only marks,mark=*,error bars/.cd,y dir=both,y explicit]
coordinates {(0.5,-0.028) +- (0,0.010) (1,0.002) += (0,0.015) -= (0,0.018) (3,0.116) +- (0,0.020)};
\addlegendentry{Three-size fit}
\addplot[red!75!black,only marks,mark=square*,error bars/.cd,y dir=both,y explicit]
coordinates {(0.5,-0.048) += (0,0.022) -= (0,0.023) (3,0.111) += (0,0.036) -= (0,0.037)};
\addlegendentry{$5887\rightarrow15435$}
\end{axis}
\end{tikzpicture}
\caption{Disorder-dependent participation contrast at identical one-site Gaussian marginals. The three-size fit and the largest adjacent-size step have the same signs at the weak and strong endpoints. Bars show the pointwise $95\%$ intervals in Tables~\ref{tab:alpha} and \ref{tab:pairs}. }
\label{fig:contrast}
\end{figure}
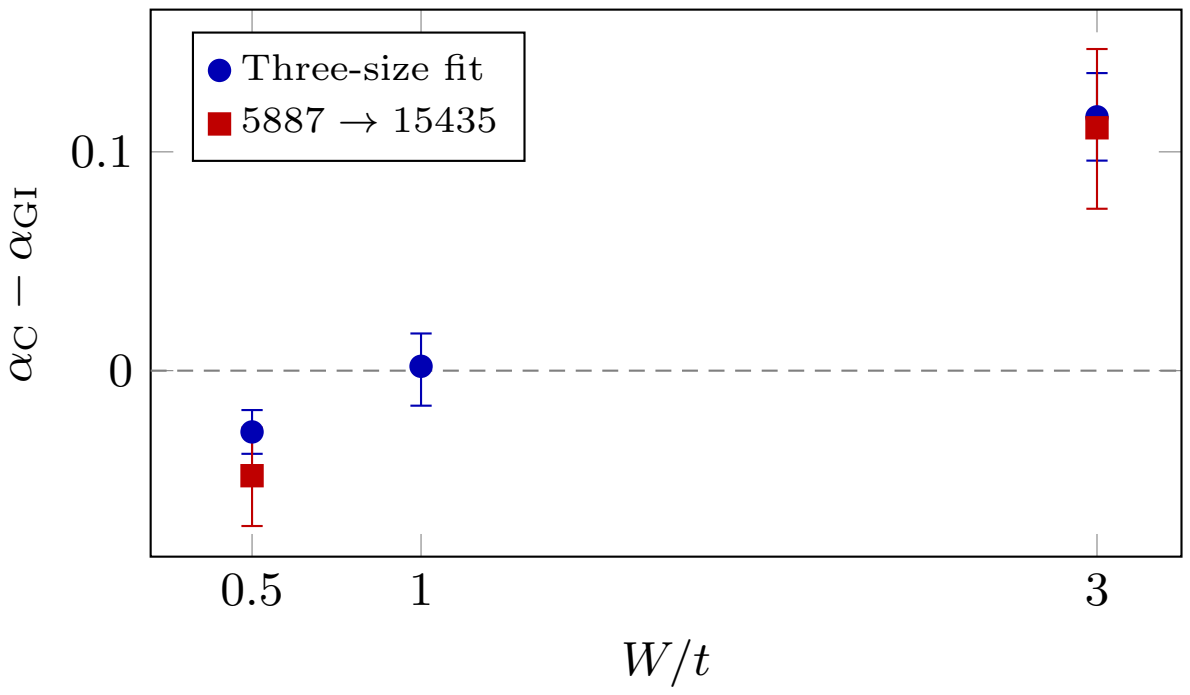

At weak disorder, the Comtet field has the smaller growth exponent of $r_N$. At strong disorder, it has the larger growth exponent. Since the local distributions agree exactly, these differences arise from the spatial organization of the Gaussian potential on the common graph. The exponent summarizes a full energy window containing states with different spatial structures.

The point at $W/t=1$ helps separate a level difference from a scaling difference. Its exponent interval includes zero, while the fixed-size statistic
\begin{equation}
\theta_N=\ln\langle I_s\rangle_\GI-\ln\langle I_s\rangle_\Cdis
\label{eq:theta}
\end{equation}
has small negative offsets, approximately $-0.03$ to $-0.01$. Two of the three sizes reject the level null in the Holm-adjusted tests. A prefactor difference in $r_N$ can therefore coexist with an unresolved slope difference. The nine-test level family is defined in Appendix~\ref{app:statistics}.

\subsection{Persistence at the largest graph}

Table~\ref{tab:pairs} resolves the adjacent-size slopes, including the extension from $5887$ to $15435$ vertices. At $W/t=0.5$, the largest two steps give negative contrasts, whereas the smallest step has an interval spanning zero. At $W/t=3$, all three reported steps give positive contrasts. The largest step reproduces the signs of the three-size fit at both endpoints (Fig.~\ref{fig:contrast}).

\begin{table}[tb]
\caption{Adjacent-size exponent contrasts, $100$ realizations per class and point. Pointwise $95\%$ intervals use $10^4$ bootstrap resamples. The step dependence displays finite-size variation separately from disorder-sampling uncertainty.}
\label{tab:pairs}
\centering
\begin{tabular}{rrrl}
\toprule
$W/t$&$N_a\to N_b$&$\Delta\alpha_{ab}$&$95\%$ CI\\
\midrule
0.5&$847\to2240$&$-0.012$&$[-0.033,+0.009]$\\
&$2240\to5887$&$-0.045$&$[-0.068,-0.023]$\\
&$5887\to15435$&$-0.048$&$[-0.071,-0.026]$\\
3&$847\to2240$&$+0.171$&$[+0.127,+0.218]$\\
&$2240\to5887$&$+0.060$&$[+0.019,+0.101]$\\
&$5887\to15435$&$+0.111$&$[+0.074,+0.147]$\\
\bottomrule
\end{tabular}
\end{table}

The strong-disorder sequence $+0.171,+0.060,+0.111$ varies nonmonotonically with graph size. The weak-disorder contrast becomes more negative across the two largest steps. In the strong-disorder channel decomposition, the GI slopes decrease from approximately $0.246$ for $2240\to5887$ to $0.199$ for $5887\to15435$, while the Comtet slope remains near $0.31$. The extension therefore strengthens the finite-graph sign comparison. The changing step magnitudes remain relevant when interpreting the fitted exponents.

\section{Higher participation moments and spectral structure}
\label{sec:moments}

Higher moments reveal spatial inhomogeneity beyond the mean IPR. Table~\ref{tab:Dq} reports $D_2,D_3,D_4$ over the three main sizes. Every class and disorder strength has $D_2>D_3>D_4$. Increasing $q$ emphasizes sites with larger wave-function amplitudes, so the decrease shows that the scaling of high-amplitude regions differs from that of the second moment. With the present averaging convention, $D_2$ is exactly the previously defined $\alpha$; the additional information is the dependence on $q>2$.

\begin{table}[tb]
\caption{Generalized participation dimensions from the three main $\{7,3\}$ sizes, normalized to $\ln N$. They are finite-size moment slopes, with $D_2=\alpha$ by construction.}
\label{tab:Dq}
\centering
\begin{tabular}{crrrr}
\toprule
Class&$W/t$&$D_2$&$D_3$&$D_4$\\
\midrule
C&0.5&0.531&0.350&0.251\\
C&1&0.493&0.334&0.243\\
C&3&0.356&0.229&0.156\\
GI&0.5&0.560&0.374&0.251\\
GI&1&0.492&0.332&0.237\\
GI&3&0.240&0.149&0.106\\
\bottomrule
\end{tabular}
\end{table}

The weak-disorder C--GI difference becomes small at $q=4$ at the displayed precision, even though it is resolved for $D_2$. At strong disorder, C has the larger value at every listed moment order. This is consistent with covariance influencing more than one feature of the amplitude distribution. Arithmetic moment averages emphasize rare large amplitudes; here their $q$ dependence characterizes inhomogeneous participation scaling over the measured sizes. The larger-size moment data cover $W/t=0.5,3$; the $W/t=1$ comparison uses the three main sizes.

The spacing ratios provide a complementary spectral diagnostic (Table~\ref{tab:spacing}). At $W/t=0.5$, both classes have mean ratios close to the large-matrix Gaussian orthogonal ensemble reference, approximately $0.5307$ \cite{Atas2013}. With increasing disorder the values move toward the Poisson reference $2\ln2-1\simeq0.3863$, while remaining above it at all reported points. At $W/t=3$, GI has the smaller spacing ratio as well as the smaller participation exponent. Thus the stronger-disorder contrast appears in both wave-function and spectral observables.

\begin{table}[tb]
\caption{Spacing-ratio means pooled from per-realization means across $N=847,2240,5887$, with bootstrap $95\%$ intervals. Each entry summarizes all three sizes.}
\label{tab:spacing}
\centering
\begin{tabular}{crl}
\toprule
Class&$W/t$&$\langle\rho\rangle$ [$95\%$ CI]\\
\midrule
C&0.5&$0.5250\ [0.5232,0.5269]$\\
C&1&$0.5219\ [0.5199,0.5239]$\\
C&3&$0.5050\ [0.5027,0.5072]$\\
GI&0.5&$0.5234\ [0.5214,0.5253]$\\
GI&1&$0.5218\ [0.5198,0.5237]$\\
GI&3&$0.4883\ [0.4860,0.4906]$\\
\bottomrule
\end{tabular}
\end{table}

The pooled statistic describes the spectral change accompanying the moment data, averaged over graph sizes and the full energy window. State-resolved classification requires an energy- and size-resolved analysis.

\section{An exponential-kernel comparison at fixed marginals}
\label{sec:expresults}

\subsection{Kernel family and normalization}

To probe which features are specific to the Comtet kernel, we construct a tunable exponential family,
\begin{align}
L_{\xi,ij}&=e^{-d_{ij}/\xi}\mathbf 1_{d_{ij}\le R_\xi},\\
Q_{\xi,i}&=\sum_jL_{\xi,ij}^2,\\
V_i^\Expdis&=\sigma_i\frac{\sum_jL_{\xi,ij}z_j}{\sqrt{Q_{\xi,i}}},
\qquad z_j\overset{\mathrm{iid}}{\sim}{\cal N}(0,1).
\label{eq:exp}
\end{align}
The $\sigma_i$ are those of the Comtet field on the same graph. The deterministic row normalization gives $\operatorname{Var}V_i^\Expdis=\sigma_i^2$ exactly. It preserves Gaussianity and requires no realization-dependent rescaling.

The baseline support includes every graph pair: $R_\xi=r_{\rm geo}=\max_{ij}d_{ij}$. Unlike the Comtet matrix, $L_\xi$ retains its diagonal, $L_{\xi,ii}=1$. Therefore $L_\xi\to I$ as $\xi\to0^+$ on a fixed finite graph, and the resulting field tends in distribution to GI. The input $\xi$ is a kernel length; its relation to the field covariance follows from the row overlaps, just as in Eq.~\eqref{eq:cov}. Changes of self-term, short-distance shape, and support mean that varying $\xi$ is not a continuous interpolation through the Comtet ensemble.

\subsection{Same-sign response over the sampled lengths}

Table~\ref{tab:exp} gives $\alpha_\Expdis-\alpha_\GI$ at $\xi/\ell=0.25,0.5,1,2,3$ and $W/t=0.5,3$. Two independent batches, A and B, each with $50$ Exp source realizations per size and state point, give positive contrasts throughout this grid. The reported pointwise intervals exclude zero in both batches, with half-widths at most $0.028$.

\begin{table}[tb]
\caption{Exponential-family exponent contrasts relative to GI. Entries are batch A/B estimates from separate sets of $50$ Exp source realizations per size and state point, fitted across the three main graph sizes.}
\label{tab:exp}
\centering
\begin{tabular}{rcc}
\toprule
$\xi/\ell$&$W/t=0.5$&$W/t=3$\\
\midrule
0.25&$+0.2473/+0.2450$&$+0.5510/+0.5500$\\
0.50&$+0.2362/+0.2355$&$+0.5352/+0.5353$\\
1.00&$+0.2070/+0.2122$&$+0.5192/+0.5307$\\
2.00&$+0.1828/+0.1802$&$+0.4923/+0.5009$\\
3.00&$+0.1763/+0.1750$&$+0.4930/+0.4892$\\
\bottomrule
\end{tabular}
\end{table}

\begin{figure}[tb]
\centering
\begin{tikzpicture}
\begin{axis}[width=8.4cm,height=5.3cm,xlabel={$\xi/\ell$},
ylabel={$\alpha_\Expdis-\alpha_\GI$},xmin=0,xmax=3.2,ymin=0,ymax=0.64,
tick label style={font=\small},label style={font=\small},
legend style={font=\scriptsize,at={(0.98,0.97)},anchor=north east},
legend cell align=left]
\addplot[blue!70!black,mark=*,thick] coordinates {(0.25,0.2450)(0.5,0.2355)(1,0.2122)(2,0.1802)(3,0.1750)};
\addlegendentry{$W/t=0.5$, batch B}
\addplot[red!75!black,mark=square*,thick] coordinates {(0.25,0.5500)(0.5,0.5353)(1,0.5307)(2,0.5009)(3,0.4892)};
\addlegendentry{$W/t=3$, batch B}
\addplot[blue!70!black,only marks,mark=o] coordinates {(0.25,0.2473)(0.5,0.2362)(1,0.2070)(2,0.1828)(3,0.1763)};
\addlegendentry{$W/t=0.5$, batch A}
\addplot[red!75!black,only marks,mark=square] coordinates {(0.25,0.5510)(0.5,0.5352)(1,0.5192)(2,0.4923)(3,0.4930)};
\addlegendentry{$W/t=3$, batch A}
\end{axis}
\end{tikzpicture}
\caption{Positive participation contrasts at the five sampled exponential-kernel lengths, using the estimates in Table~\ref{tab:exp}. Symbols show point estimates; lines guide the eye between batch B values, and batch A uses open symbols.}
\label{fig:exp}
\end{figure}
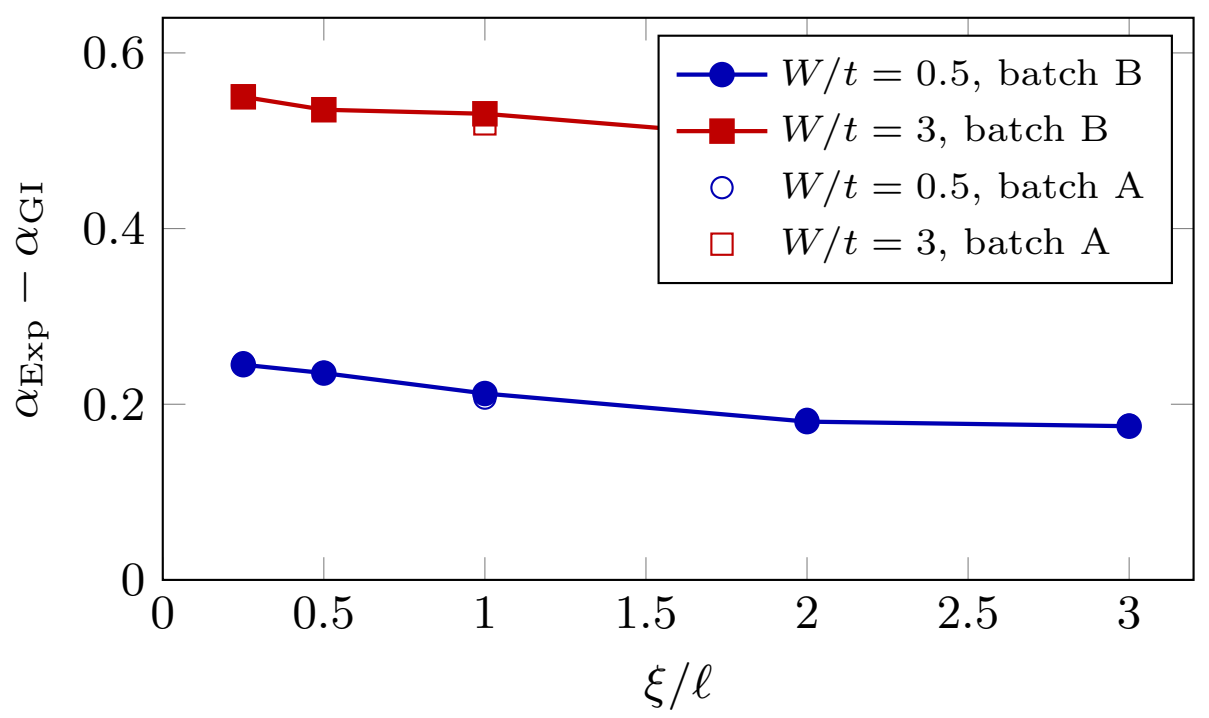

The sampled exponential family has a positive response at weak disorder, in contrast to the negative Comtet response. The positive values persist across the tested lengths, including $\xi=0.25\ell$, while the exact $\xi\to0^+$ limit is GI. Thus this finite grid resolves a same-sign response over its sampled range; it does not determine the behavior at intermediate or smaller lengths.

\subsection{Support and continuum tail scales}

For an infinite continuum exponential kernel, the radial squared-kernel tail scales as
\begin{equation}
e^{-2r/\xi}\sinh(r/\ell)
\sim\frac{1}{2}e^{(1/\ell-2/\xi)r}.
\label{eq:exptail}
\end{equation}
Thus $\xi=2\ell$ separates convergent and divergent infinite-volume radial integrals. Every finite-graph sum in Eq.~\eqref{eq:exp} remains finite. The continuum integrability threshold provides a reference for support sensitivity of the finite normalized field.

The explicit support test uses the $N=847$ graph and the same Gaussian source realizations for each variant. Its diameter is $9.77\ell$, so imposing $R_\xi=12\ell$ leaves all pairs intact. A cutoff at $4.9\ell$ changes the mean IPR by up to $1.59\%$ at $W/t=0.5$ and $4.88\%$ at $W/t=3$ for $\xi=3\ell$. The full table appears in Appendix~\ref{app:support}. This comparison measures the single-size mean-IPR sensitivity. It establishes that long input kernels retain measurable sensitivity to the finite support used in their construction.
\section{Discussion}
\label{sec:discussion}

\subsection{A concentrated variance budget can retain a covariance response}

The central distinction is between a source-distance budget and a wave-function response. The former is a normalized sum of $K_{ij}^2$ and is dominated by nearby sources. The latter is measured after the random field enters a Hamiltonian, mixes eigenstates, and changes their spatial amplitudes. The field covariance is $a_W^2KK^T$, so even a concentrated source-distance budget leaves a structured matrix of correlations between vertices.

Exact marginal matching makes this distinction operational. A comparison with uniform box disorder would also change the shape and spatial profile of the local distributions. The GI ensemble preserves both. The opposite signs at the weak and strong endpoints therefore express a response to the joint spatial statistics of a Gaussian field with a common boundary-dependent variance profile.

Covariance changes the disorder matrix elements in the clean eigenbasis, suggesting a state-dependent response to the potential's spatial structure. Resolving this mechanism would require energy- and state-resolved correlations between those matrix elements and participation. The measured sign contrast provides the controlled ensemble comparison for such an analysis.

\subsection{The kernel family separates decay from short-distance structure}

The exponential family gives a second way to organize the problem. It holds the same local Gaussian distributions fixed while changing the overlap geometry of the kernel rows. Across the sampled lengths its participation contrast is positive at both disorder strengths, whereas Comtet disorder has a negative weak-disorder contrast. Thus the asymptotic decay of a kernel, by itself, does not specify the sign of the participation response.

The comparison also clarifies why the input length is insufficient as a common coordinate. Comtet disorder excludes self-coupling and has a logarithmic near-source form; the exponential family retains self-coupling and uses deterministic row normalization on full graph support. These choices affect the normalized covariance at short distances even when far tails have comparable inverse lengths. Their physical role is therefore encoded in the full covariance matrix, rather than in a fitted tail mass alone. The tail identities in Appendix~\ref{app:tail} make this point explicit: recovering the input exponential decay from a shell-slope difference is a construction identity whose observable content is separate from the participation contrast.

\subsection{Finite open geometry and the meaning of scaling}

Finite-size scaling and its corrections are central to the interpretation of localization observables \cite{Thouless1974,KramerMacKinnon1993,SlevinOhtsuki1999}. Tree and random-regular-graph models also expose the importance of graph growth and connectivity \cite{AbouChacra1973,Tikhonov2016}; a regular hyperbolic tiling additionally has short loops and, in the present construction, an explicit open boundary. The graph's boundary participates in both the variance measure and the spectrum. Matching local variances retains its effect on the one-site distribution, but a common boundary does not guarantee that its coupling to two different covariances cancels. The reported $\alpha$ and $D_q$ are therefore effective exponents for complete windows on the given open graph sequence. Their normalization by $\ln N$ is useful for comparing ensembles on precisely that sequence.

Two types of size dependence are resolved. The squared-kernel radius changes little across the larger graphs, while adjacent participation slopes continue to vary. A stable input-variance quantile therefore coexists with a size-dependent spectral response. Bootstrap intervals quantify disorder-sampling uncertainty at fixed sizes, while adjacent slopes expose finite-size variation. Pooled spacing ratios and arithmetic participation moments describe complementary features of this window. Establishing a mobility edge or a critical multifractal spectrum would require resolving energy dependence and asymptotic size scaling.

The present observable-based comparison complements continuum self-consistent localization theories \cite{VollhardtWoelfle1980} within the broader study of disorder-driven electronic transitions \cite{BelitzKirkpatrick1994}. Spectral-density calculations on hyperbolic graphs \cite{mosseri2023Density} supply another complementary view; polynomial-expansion methods can access such spectral densities \cite{Weisse2006}, whereas the participation moments here require the window eigenstates themselves.

Compact hyperbolic quotients offer a geometrically different comparison \cite{MaciejkoRayan2021,Maciejko2022}. Gauss--Bonnet requires genus at least two for a closed orientable surface of constant negative curvature and positive area. An ordinary flat torus cannot implement the same geometry by identifying opposite edges of the present disk. A quotient requires new vertex and edge identifications, geodesic distances, and disorder kernels (Appendix~\ref{app:boundary}). It is a distinct model construction with its own finite-size sequence.

The results provide a practical reference for correlated scalar disorder in open hyperbolic systems: retain the full local distribution, specify the covariance independently, and measure participation and spectral structure alongside the kernel's variance budget. The field definitions specify local detunings and their covariance separately, providing a formulation for studying correlated disorder in programmable circuit and photonic graphs.
\section{Conclusion}

We have separated the geometric variance budget of Comtet disorder from its covariance-dependent wave-function response on open hyperbolic lattices. The squared kernel is concentrated within a few curvature lengths. Its size-resolved $95\%$ radius is close to $2.45\ell$ along the largest $\{7,3\}$ graphs, and the discrete shell weights resolve the geometry associated with its separation from the cutoff-matched continuum reference.

At exactly matched one-site Gaussian distributions, the participation-exponent contrast is negative at weak disorder and positive at strong disorder. Both signs persist in the extension to $15435$ vertices. Higher participation moments and spacing ratios add information about the spatial amplitudes and spectral correlations accompanying this response. A row-normalized exponential-kernel family gives positive contrasts at all sampled lengths at $W/t=0.5,3$, showing that the tested covariance constructions give distinct responses beyond their far-tail decay.

Together, these results establish a finite-graph comparison in which local disorder distributions, source-distance variance, and wave-function participation can be resolved independently. The distinction gives a concrete way to study how negative-curvature geometry organizes correlated disorder and its spectral consequences.
\begin{acknowledgments}
This work was supported by the Science Research Foundation for High-Level Talents of Anhui University of Science and Technology under Grant YJ20240002 and by the National Natural Science Foundation of China under Grant No.~11304318.
\end{acknowledgments}
\appendix
\section{Electrostatic Green function and continuum variance}
\label{app:green}

For curvature $-1/\ell^2$, the geodesic polar metric is
\begin{equation}
ds^2=dr^2+\ell^2\sinh^2(r/\ell)\,d\theta^2.
\end{equation}
The area element is $dA=\ell\sinh(r/\ell)\,dr\,d\theta$. Acting on a radial function, the Laplace--Beltrami operator is
\begin{equation}
\Delta G=G''(r)+\ell^{-1}\coth(r/\ell)G'(r).
\end{equation}
Away from the source, $-\Delta G=0$, and multiplication by $\sinh(r/\ell)$ gives
\begin{equation}
\frac{d}{dr}\left[\sinh(r/\ell)G'(r)\right]=0.
\end{equation}
Consequently $G'(r)=C/\sinh(r/\ell)$. The flux through a circle of radius $r$ must satisfy
\begin{equation}
-2\pi\ell\sinh(r/\ell)G'(r)=1,
\end{equation}
which fixes $C=-1/(2\pi\ell)$. Integration uses
\begin{equation}
\int\frac{dr}{\ell\sinh(r/\ell)}
=\ln\tanh\frac{r}{2\ell}.
\end{equation}
Imposing $G(\infty)=0$ then gives Eq.~\eqref{eq:Phi}. Its short-distance and far-distance limits are
\begin{align}
\Phi_H(r)&=-\frac{1}{2\pi}\ln\frac{r}{2\ell}
+O(r^2/\ell^2),\\
\Phi_H(r)&=\frac{1}{\pi}\left[
e^{-r/\ell}+\frac{1}{3}e^{-3r/\ell}+\cdots\right].
\end{align}
The source singularity is integrable after squaring and multiplying by the radial measure; the lattice construction excludes the self-term explicitly.

To evaluate the continuum radial integral, set $x=r/\ell$ and $u=\tanh(x/2)$. Since
\begin{equation}
\sinh x\,dx=\frac{4u}{(1-u^2)^2}\,du,\qquad
\ln\coth(x/2)=-\ln u,
\end{equation}
we obtain
\begin{align}
{\cal I}&=\int_0^\infty\Phi_H^2(r)\sinh(r/\ell)\,dr\\
&=\frac{\ell}{\pi^2}\int_0^1
\frac{u\ln^2u}{(1-u^2)^2}\,du.
\end{align}
Expanding $(1-u^2)^{-2}=\sum_{m=1}^\infty m\,u^{2m-2}$ yields
\begin{align}
{\cal I}&=\frac{\ell}{\pi^2}\sum_{m=1}^\infty
m\int_0^1u^{2m-1}\ln^2u\,du\\
&=\frac{\ell}{\pi^2}\sum_{m=1}^\infty
\frac{1}{4m^2}
=\frac{\ell}{24}.
\end{align}
The integral identity follows by differentiating
$\int_0^1u^{a-1}du=1/a$ twice with respect to $a$.
All terms in the series integral are nonnegative, so monotone convergence justifies interchanging the sum and integral. Multiplying by the angular area factor gives $\int_{\HH^2}\Phi_H^2\,dA=\pi\ell^2/12$. A source density would enter a continuum field variance as an additional factor. This makes explicit why the radial number $\ell/24$ cannot directly replace a graph's $S_2$.

\section{Shell controls and the continuum--discrete offset}
\label{app:offset}

Table~\ref{tab:distinct} lists selected distinct-distance shells. The first three are consecutive, while the last two show the approach to the $95\%$ quantile; intervening shells are included in the cumulative column. These exact-distance groups differ from the broad intervals of Table~\ref{tab:shell} and from the histogram used to interpolate $r_{95}$.

\begin{table}[tb]
\caption{Selected shells on $\{7,3\}$ with $N=847$. Fractions refer to the full normalized squared-kernel weight. The last cumulative value is below $0.95$.}
\label{tab:distinct}
\centering
\begin{tabular}{rrr}
\toprule
$r/\ell$&Shell fraction&Cumulative fraction\\
\midrule
0.5663&0.3901&0.3901\\
0.9928&0.2339&0.6239\\
1.2136&0.1330&0.7569\\
2.1916&0.0240&0.9379\\
2.3319&0.0116&0.9495\\
\bottomrule
\end{tabular}
\end{table}

The control radii appear in Table~\ref{tab:offset}. The hard-cut continuum comparison uses the same nominal $6\ell$ kernel cutoff. For the reverse comparison, the discrete profile is extended by a continuum tail. Let $p_{\rm tail}=1-F_\infty^{\rm ct}(R_C)$ and define
\begin{equation}
\begin{split}
F_{\rm join}(r)&=(1-p_{\rm tail})F_N(r)+p_{\rm tail}G(r),\\
G(r)&=\begin{cases}
0,&r\le R_C,\\
\dfrac{F_\infty^{\rm ct}(r)-F_\infty^{\rm ct}(R_C)}{p_{\rm tail}},&r>R_C.
\end{cases}
\end{split}
\end{equation}
The joined-profile radii are quantiles of this normalized extension. The hard-cut reference and the joined profile answer different questions: the former compares a truncated continuum profile with the discrete graph, while the latter compares a tail-extended discrete profile with the infinite-support continuum.

\begin{table}[tb]
\caption{Discrete radii and continuum control radii. Lengths are in units of $\ell$. Here $r_{\rm geo}$ is the graph diameter; the hard-cut continuum radius is $3.1228\ell$. Joined-profile entries exist only for the four $\{7,3\}$ graphs.}
\label{tab:offset}
\centering
\begin{tabular}{ccrrrr}
\toprule
Tiling&$n$&$r_{95}^{\rm disc}$&$r_{\rm geo}$&Hard cut&Joined\\
\midrule
$\{7,3\}$&5&2.4302&9.77&3.1228&2.4585\\
&6&2.4432&11.90&3.1228&2.4711\\
&7&2.4481&14.04&3.1228&2.6256\\
&8&2.4499&16.17&3.1228&2.6269\\
$\{8,3\}$&5&2.5509&13.89&3.1228&---\\
&6&2.5524&16.94&3.1228&---\\
\bottomrule
\end{tabular}
\end{table}

The joined profiles use $p_{\rm tail}=2.9\times10^{-3}$ and joining radius $6\ell$. Their quantiles remain below the infinite-support continuum value $3.1792\ell$; the hard-cut continuum value likewise remains above the discrete radii. These comparisons test the specified profile modifications. A quadrature-based estimate, using $f(r)=\Phi_H^2(r)\sinh(r/\ell)$, takes the form
\begin{equation}
\delta_{\rm an}=-\frac{h^2}{12}
\left.\frac{d\ln f(r)}{dr}\right|_{r_{95}}
=2.08\times10^{-4}\ell.
\end{equation}
Its scale is much smaller than the observed offsets. This leading term concerns a smooth radial quadrature; endpoint and quantile-interpolation terms, together with the discrete shell populations, enter the full comparison.

\section{Cross-tiling variance survey}
\label{app:survey}

Table~\ref{tab:survey} compares one graph size per tiling. Its mean and population standard deviation are $2.72\ell$ and $0.29\ell$. This survey uses a quantile convention different from the fixed-estimator size sequences of Table~\ref{tab:r95}; its spread measures variation across geometries.

\begin{table}[tb]
\caption{Cross-tiling survey of the $95\%$ variance radius. These values use the survey's quantile convention; Table~\ref{tab:r95} uses the fixed-bin size-resolved estimator.}
\label{tab:survey}
\centering
\begin{tabular}{crr}
\toprule
Tiling&$N$&$r_{95}/\ell$\\
\midrule
$\{4,5\}$&384&2.61\\
$\{5,4\}$&625&2.61\\
$\{6,4\}$&1176&3.09\\
$\{7,3\}$&5887&2.48\\
$\{8,3\}$&2888&2.56\\
$\{10,3\}$&16810&2.36\\
$\{5,5\}$&2205&3.28\\
$\{3,7\}$&135&2.73\\
\bottomrule
\end{tabular}
\end{table}

\section{Statistical and numerical details}
\label{app:statistics}

The disorder realization is the resampling unit. For each class and size, the bootstrap resamples the $S$ values of $I_s$, reconstructs $r_N$, and refits the same size sequence. This retains the distinction between within-realization state averaging and between-realization weighting. The quoted intervals in Table~\ref{tab:alpha} are pointwise $95\%$ percentile intervals from $300$ resamples; a $10^4$-resample recomputation reports endpoint changes within $10^{-3}$. The pairwise intervals use $10^4$ resamples.

The separate fixed-size level tests use $\theta_N$ of Eq.~\eqref{eq:theta}, a studentized bootstrap with $10^4$ resamples, and first-order delta-method standard errors for the log-mean contrast. Holm correction \cite{Holm1979} is applied to the nine $(n,W)$ tests. Eight reject the level null, with $n=6$, $W/t=1$ the exception. These adjusted level tests and the pointwise exponent intervals concern different statistics.

An empirical permutation calibration of the nine-test procedure reports a family rejection fraction $0.049$ over $1000$ repetitions, with Clopper--Pearson interval $[0.0365,0.0643]$. A separate mean-matched empirical null gives zero rejections in $300$ repetitions and interval $[0,0.0122]$. These are calibrations of specified resampling procedures on the sample distributions. Their uncertainty concerns the resampling calibration at the sampled graph sizes.

For the exponential family, batches A and B use sample indices $s=0$--$49$ and $s=100$--$149$. The source stream is specified by
\begin{equation}
\texttt{SeedSequence([20261010,7,3,n,6,j,s,9001])},
\end{equation}
where $j$ indexes $\xi$. Distinct $\xi$ use distinct streams. The two disorder strengths share the underlying $z$ at fixed indices, and support variants share $z$ within each comparison. The sampled strengths and support variants therefore have explicit dependence. In the empirical GI-null calibration, pseudo-Exp batches are drawn independently with replacement from the $100$ stored GI realizations at each size and disorder strength; the full GI set supplies the reference. The event is selection of an opposite-sign candidate across neighboring $\xi$ values followed by confirmation in a separate pseudo-Exp batch. Two calibration batches, each with $1000$ Monte Carlo trials, give zero such events and a Wilson $95\%$ upper bound \cite{Wilson1927} of $0.0038$ per batch. This bound quantifies Monte Carlo trial uncertainty conditional on the fixed empirical GI distribution; population-level calibration and detection power require separate evaluation.

\subsection{Solver, cutoff, and stored quantities}
\label{app:numerics}

The same-window dense/Lanczos comparison reports eigenpair residuals at most $3.2\times10^{-14}$, norm errors at most $1.4\times10^{-14}$, and relative mean-IPR differences at most $1.3\times10^{-12}$. Completeness of the requested window is part of the production protocol. At the largest graph, the calculation uses shift-invert Lanczos with a returned population extending beyond the target window; overlapping dense comparisons report mean-IPR agreement below $10^{-12}$. These checks concern solver agreement for the tested windows.

A Comtet cutoff benchmark reports a squared-kernel tail share $2.4\times10^{-4}$ beyond $6\ell$ and mean participation changes at most $1.3\%$ between cutoffs $6\ell$, $8\ell$, and full range. A separate scan uses $50$ realizations at $N=847$ and $20$ at each of $2240,5887$, for each class and configuration. It varies both the cutoff and whether the normalization is recomputed. Weak- and strong-endpoint point estimates retain their negative and positive signs, respectively. Some strong-disorder intervals for the larger pair include zero in this lower-sample scan. The largest difference of exponent contrasts between the two normalizations at a common cutoff is $7.14\times10^{-4}$. This is a tail-cutoff test beyond $6\ell$, not a test that truncating at the $95\%$ variance radius leaves the spectrum invariant.

Production kernels are cached in single precision. A double-precision comparison reports relative changes of $2.2\times10^{-8}$ in $S_2$ and $1.4\times10^{-7}$ in the main participation ratios. Per-realization records retain window populations, mean IPRs, potential moments, solver residuals, and random seeds. The C--GI source stream is
\begin{equation}
\texttt{SeedSequence([20261010,7,3,n,c,s,9001])},
\end{equation}
with $c=2$ for C and $c=5$ for GI; the starting-vector stream appends $123456$. Different class codes give independent samples. The numerical implementation uses NumPy and SciPy \cite{NumPy2020,SciPy2020}. The kernel-decomposition figure uses Matplotlib \cite{Matplotlib2007}; the participation-contrast figures are drawn with PGFPlots from the tabulated estimates.

\section{Exponential-kernel support sensitivity}
\label{app:support}

Table~\ref{tab:support} gives all seven single-size support comparisons used here. The observable is the average of per-realization window IPRs on $N=847$, using $50$ shared source realizations. Percentage changes are computed before rounding the displayed mean IPRs. Since the graph diameter is $9.77\ell$, the $12\ell$ variant is identical to full graph support by construction. The $4.9\ell$ variant removes graph pairs and shows the larger effects at the longer tested lengths.

\begin{table*}[tb]
\caption{Support sensitivity of the exponential field on $N=847$. $\overline I$ denotes the dimensionless mean window IPR. The two percentage columns are changes relative to full graph support. This table does not compare fitted exponents.}
\label{tab:support}
\centering
\begin{tabular}{rrrrrrr}
\toprule
$W/t$&$\xi/\ell$&Full support $\overline I$&$R_\xi=12\ell$&$R_\xi=4.9\ell$&
Change ($12\ell$)&Change ($4.9\ell$)\\
\midrule
0.5&0.25&0.003462&0.003462&0.003462&0.00\%&0.00\%\\
0.5&1&0.003595&0.003595&0.003595&0.00\%&$+0.02\%$\\
0.5&2&0.003627&0.003627&0.003650&0.00\%&$+0.62\%$\\
0.5&3&0.003584&0.003584&0.003641&0.00\%&$+1.59\%$\\
3&1&0.004248&0.004248&0.004254&0.00\%&$+0.15\%$\\
3&2&0.004281&0.004281&0.004366&0.00\%&$+1.98\%$\\
3&3&0.004150&0.004150&0.004353&0.00\%&$+4.88\%$\\
\bottomrule
\end{tabular}
\end{table*}

Finite-sample construction diagnostics compare the generated field with its analytic covariance. Variance and covariance comparisons satisfy their specified tolerances in $20/20$ cells each; the covariance diagnostic has absolute standardized-error medians $0.437$--$0.832$ and $95$th percentiles $1.18$--$2.60$. Correlation-length ordering is consistent with the input-length ordering in $4/4$ groups when crossings outside the measured distance range are treated as censored. A separate $50$-replicate check of the variance-error estimate satisfies its tolerance in $29/30$ cells; the exception is $n=5$, $W/t=3$, $\xi=2\ell$. These diagnostics concern sampled-field construction, while participation scaling is measured independently.

\section{Tail-slope diagnostics and their interpretation}
\label{app:tail}

A shell-slope diagnostic separates the kernel contribution from the graph pair measure. Nonzero off-diagonal pairs are grouped by distances rounded to $10^{-6}\ell$; both logarithmic profiles are fitted by ordinary least squares with an intercept and equal shell weights. The interval starts at the squared-kernel $r_{95}$ and ends at the largest distance with nonzero kernel weight. Both slopes must satisfy $R^2>0.98$ and slope standard error below $0.1\ell^{-1}$ to describe a common exponential tail. Let $v_k$ be the number of ordered pairs in a distinct-distance shell and $w_k$ its squared-kernel weight. For the exponential kernel,
\begin{equation}
w_k=v_ke^{-2r_k/\xi},
\qquad
\ln w_k=\ln v_k-\frac{2r_k}{\xi}.
\label{eq:tailidentity}
\end{equation}
If $a$ and $b$ are the fitted slopes of $\ln w_k$ and $\ln v_k$ versus $r_k$, the linearity of a fit with identical points, weights, and intercept convention gives
\begin{equation}
a=b-\frac{2}{\xi},\qquad
\mu_{\rm eff}\equiv\frac{b-a}{2}=\frac{1}{\xi}.
\end{equation}
Distance grouping to finite precision modifies the identity at that precision. The close recovery of $1/\xi$ is therefore chiefly a consequence of the construction. It is useful as an implementation consistency check, while the wave-function comparison supplies different information.

\begin{table*}[tb]
\caption{Comtet tail diagnostics. Slopes and $\mu_{\rm eff}$ are in $\ell^{-1}$; SE is the standard error of the tail slope. All six rows fail the joint exponential-fit quality criterion defined in the text. Here $\kappa_{\rm graph}$ denotes the all-pair shell-count slope, not an infinite-plane radial volume exponent.}
\label{tab:tailC}
\centering
\begin{tabular}{ccrrrrrrc}
\toprule
Tiling&$n$&$a$&$R_a^2$&SE&$\kappa_{\rm graph}$&$R_\kappa^2$&$\mu_{\rm eff}$&Joint fit\\
\midrule
$\{7,3\}$&5&$-2.3977$&0.944&0.045&$-0.3972$&0.317&1.0002&No\\
&6&$-2.3493$&0.943&0.044&$-0.3488$&0.269&1.0002&No\\
&7&$-2.3322$&0.943&0.044&$-0.3318$&0.251&1.0002&No\\
&8&$-2.3259$&0.943&0.044&$-0.3254$&0.244&1.0002&No\\
$\{8,3\}$&5&$-2.2721$&0.943&0.055&$-0.2717$&0.192&1.0002&No\\
&6&$-2.2635$&0.943&0.055&$-0.2631$&0.183&1.0002&No\\
\bottomrule
\end{tabular}
\end{table*}

\begin{table*}[tb]
\caption{Exponential-kernel tail fits over the specified tail intervals. All slopes and SE are in $\ell^{-1}$. The values of $\mu_{\rm eff}$ reflect Eq.~\eqref{eq:tailidentity}; the joint exponential-fit quality criterion is not satisfied in these rows.}
\label{tab:tailE}
\centering
\begin{tabular}{rrrrrrr}
\toprule
$n$&$\xi/\ell$&$a$&$R_a^2$&SE&$\kappa_{\rm graph}$&$\mu_{\rm eff}$\\
\midrule
5&0.25&$-8.8225$&0.996&0.0098&$-0.8225$&4.0000\\
5&0.50&$-4.8285$&0.988&0.0098&$-0.8285$&2.0000\\
5&1.00&$-2.8501$&0.965&0.0100&$-0.8501$&1.0000\\
5&2.00&$-1.9758$&0.901&0.0123&$-0.9758$&0.5000\\
5&3.00&$-1.8311$&0.766&0.0205&$-1.1645$&0.3333\\
6&0.25&$-8.8660$&0.996&0.0044&$-0.8660$&4.0000\\
6&1.00&$-2.8759$&0.962&0.0044&$-0.8759$&1.0000\\
6&3.00&$-1.7814$&0.803&0.0071&$-1.1147$&0.3333\\
7&0.25&$-8.9051$&0.995&0.0019&$-0.9051$&4.0000\\
7&1.00&$-2.9079$&0.957&0.0019&$-0.9079$&1.0000\\
7&3.00&$-1.7454$&0.820&0.0026&$-1.0787$&0.3333\\
\bottomrule
\end{tabular}
\end{table*}

Tables~\ref{tab:tailC} and \ref{tab:tailE} give the supporting fits, including those with inadequate joint fit quality. The negative $\kappa_{\rm graph}$ values describe the selected tail of the finite graph's all-pair shell histogram. This all-pair tail measure differs from shells about one point in an infinite homogeneous hyperbolic plane, whose radial growth rate is positive $1/\ell$. For Comtet disorder, the value near $1/\ell$ likewise reflects the decay of the known input kernel over the fit interval. Their interpretation is therefore tied to the input kernel and finite graph's pair measure.

\section{Compact hyperbolic boundaries and annular geometries}
\label{app:boundary}

Gauss--Bonnet gives a direct constraint on a compact orientable surface without boundary:
\begin{equation}
\int_{\cal M}K\,dA=2\pi\chi({\cal M}),\qquad
\chi({\cal M})=2-2g.
\end{equation}
At constant curvature $K=-1/\ell^2$ this becomes
\begin{equation}
-\frac{A}{\ell^2}=4\pi(1-g),
\qquad
A=4\pi\ell^2(g-1).
\end{equation}
Positive area requires $g\ge2$. In particular, a genus-one flat torus cannot serve as a closed constant-negative-curvature periodic realization of the present model. Higher-genus hyperbolic quotients are compatible with this constraint.

For a surface with boundary, the relation contains boundary geodesic curvature and, when present, corner terms:
\begin{equation}
\int_{\cal M}K\,dA+\int_{\partial{\cal M}}k_g\,ds
+\sum_a\vartheta_a=2\pi\chi({\cal M}).
\end{equation}
An annulus therefore is not excluded by the closed-surface argument. Its boundary geometry must be specified, and it is not the same object as a closed genus-one quotient.

To construct a quotient graph, one must choose vertex and edge identifications consistent with a discrete group action. Distances on that quotient and the corresponding electrostatic kernel then have to be defined anew. These choices change the geometry entering Eq.~\eqref{eq:distance} and the rows entering Eq.~\eqref{eq:cov}. The open-graph results in this article consequently refer to their explicit boundary construction; a compact comparison requires its own model and size sequence.
\bibliography{ref}
\end{document}